\documentclass[preprintnumbers,twocolumn,amsmath,amssymb,groupedaddress,superscriptaddress]{revtex4}
\usepackage{graphicx}
\usepackage{dcolumn}
\usepackage{bm}
\usepackage[dvipsnames]{xcolor}

\begin{document}

\title{Bulk and decay properties of neutron-deficient odd-mass Hg isotopes around A = 185}

\author{O. Moreno}
\email{osmoreno@ucm.es}
\affiliation{Departamento de Estructura de la Materia, F\'isica T\'ermica y Electr\'onica, Grupo de F\'isica Nuclear, and IPARCOS, Universidad Complutense de Madrid, Av. Complutense s/n, E-28040 Madrid, Spain}
\author{P. Sarriguren}
\affiliation{Instituto de Estructura de la Materia (IEM), CSIC, Serrano 123, E-28006 Madrid, Spain}
\author{A. Algora}
\affiliation{Instituto de F\'isica Corpuscular, CSIC - Universidad de Valencia, E-46071 Valencia, Spain}
\affiliation{Institute of Nuclear Research (ATOMKI), P. O. Box 51, H-4001 Debrecen, Hungary}
\author{L. M. Fraile}
\affiliation{Departamento de Estructura de la Materia, F\'isica T\'ermica y Electr\'onica, Grupo de F\'isica Nuclear, and IPARCOS, Universidad Complutense de Madrid, Av. Complutense s/n, E-28040 Madrid, Spain}
\author{S. E. A. Orrigo}
\affiliation{Instituto de F\'isica Corpuscular, CSIC - Universidad de Valencia, E-46071 Valencia, Spain}

\date{\today}

\begin{abstract}

Ground and isomeric states of the neutron-deficient odd-$A$ isotopes $^{183}$Hg, $^{185}$Hg, and $^{187}$Hg are described from a microscopic calculation based on a self-consistent, axially-deformed Hartree-Fock mean field with the Skyrme functional and pairing within BCS approximation. For each equilibrium shape and different odd-neutron states, results on mean square charge radii and magnetic dipole moments are given and analyzed in the context of their sensitivity to the nuclear deformation and to the spin and parity. Spin-isospin correlations within proton-neutron quasiparticle random phase approximation are then introduced in the nuclear states to obtain the distributions of Gamow-Teller strength and the $\beta^+/EC$ half-lives of these isotopes, whose measurements are planned at ISOLDE-CERN using total absorption gamma-ray spectroscopy techniques.

\end{abstract}

\maketitle

\section{Introduction}\label{introduction}

The nuclear phenomena of shape transition and coexistence, once believed exotic features of particular nuclei, are currently known to be much more common across the nuclear chart \cite{heyde11,wood16}. One of the most relevant regions is that of neutron-deficient isotopes around $Z=82$, where isotopic shift experiments on Hg ($Z=80$) carried out in the 70s and 80s \cite{dab79,bonn72,ulm86} revealed a sudden change in the nuclear charge radii between $^{185}$Hg and $^{186}$Hg. This jump was explained as a transition from a prolate to a slightly oblate shape \cite{frauendorf75}, the latter being also found in the ground states of the even-even Hg isotopes. The staggering of the charge radii found in Hg nuclei has been a unique feature to these isotopes until very recently \cite{bar21}. Later on, shape coexistence was observed within a very small range of excitation energy (below 1 MeV) in neutron-deficient even-even Hg and Pb isotopes \cite{julin01,andreyev00}. In this work we focus on the shape coexistence in the odd-$A$ isotopes $^{183}$Hg, $^{185}$Hg, and $^{187}$Hg and study the signatures of deformation and of spin and parity in the bulk and $\beta^+$-decay properties of their ground and isomeric states, some of them to be measured soon in a dedicated experiment at ISOLDE-CERN \cite{is707}.

The ground states of even-even Hg isotopes from $A=182$ to $A=188$ have a slight deformation, assumed predominantly oblate, and $0^+$ excited states that have been interpreted as rotational band heads of a prolate configuration \cite{julin01}. Coulomb-excitation experiments \cite{bree14} have provided information on electromagnetic properties, such as E2 transition strengths, that confirms the existence of two different shapes in these isotopes, mixed at low excitation energies but purer for higher spins.

In even-even nuclei, different 0$^+$ states very close in energy can be interpreted within a shell-model approach as multi-particle-hole excitations caused by pairing and quadrupole interactions between nucleons outside the core, which result in different deformed configurations. Within a mean-field approach, different minima in the energy-deformation curves correspond to coexisting equilibrium shapes, each matching a 0$^+$ state whose energy can be obtained through a calculation that minimizes the Hartree-Fock energy. Self-consistent, mean-field calculations in the Hg mass region of interest here have been performed in the past using Skyrme \cite{bender04,yao13,sarri05,moreno06,boillos15}, Gogny \cite{delaroche,libert,egi04,rod04,rayner10,web_gogny} or relativistic \cite{shar92,yosh94,nik02,nabi22} energy density functionals. Beyond-mean-field methods that include particle-number and angular-momentum projections and configuration mixing from a generator coordinate approach have also been considered in some of these works. The interacting boson model with configuration mixing has been recently applied as well \cite{nomura13,gramos14hg}. Triaxial deformation, which may play a role in some isotopes in this region and which has been considered in some of the previous works, does not seem to modify essentially the axially-deformed picture in the Hg isotopes under study.

More recently, a computationally-demanding Monte Carlo shell model calculation has been carried out \cite{marsh18} involving a large model space of 30 protons and up to 24 neutrons upon a doubly-magic $^{132}$Sn inert core. The mean square (m.s.) charge radius staggering, interpreted as shape alternation, is explained microscopically from a competition between the pairing correlations in the even-even isotopes, which favor sphericity, and the quadrupole and monopole components of the nucleon-nucleon interaction in the even-odd isotopes, inducing additional quadrupole deformation. The phenomenon is interpreted as a series of quantum phase transitions, with the number of neutrons being the control parameter.

In previous works \cite{sarri05,moreno06} we found that the $\beta$-decay properties in this mass region show signatures of nuclear deformation that persist under changes in the Skyrme and pairing interactions. This is particularly true for their Gamow-Teller (GT) strength distributions, which have already been measured in $^{186}$Hg \cite{algora21} using total absorption gamma-ray spectroscopy (TAGS) techniques \cite{rubio17}. The positron emission ($\beta^+$) and the electron capture ($EC$) half-lives of some ground and isomeric states of the Hg isotopes in this region have also been measured in the past few years \cite{grahn09,scheck10,gaffney14}. 

Independent access to the $\beta^+/EC$ decay of ground and isomeric states of odd-$A$ Hg isotopes for total absorption spectroscopy will provide a unique opportunity to investigate for the first time the $\beta$-decay of different shape isomers in the same nucleus. This can be achieved at ISOLDE-CERN through isomer-selective ionization using the resonance ionization laser ion source (RILIS) in a forced electron-beam-induced arc discharge ion source \cite{good16}, which can be coupled to a molten target for the production of Hg beams as proposed in \cite{is707}. Full potential of such approach can be obtained by combining the total absorption measurement of the $\beta^+/EC$ decay of isomerically-purified Hg nuclear states with a recently developed method of analysis that takes into account the penetration and summing of X-rays in the spectrometer \cite{algora21}.

In this paper we describe first the theoretical formalism used in this work (Section~\ref{theory}), followed by our results on the isotopes under study concerning bulk properties: energy-deformation curves and equilibrium shapes, charge radii, and magnetic dipole moments (Section~\ref{bulk}), as well as $\beta$-decay properties: GT strength distributions and half-lives (Section~\ref{decay}). We finish with the main conclusions of our work (Section~\ref{conclusions}).

\section{Theoretical formalism}\label{theory}

Our self-consistent microscopic description of the nuclei under study is based on an axially-deformed Hartree-Fock (HF) mean field using effective two-body, density-dependent Skyrme nucleon-nucleon interactions. The SLy4 Skyrme parametrization \cite{sly4} has been predominantly used in this work, since it has been successfully applied in a wide variety of calculations \cite{stoitsov03}. Pairing correlations between like nucleons are introduced within the BCS approximation using fixed pairing energy gaps determined phenomenologically from experimental mass differences \cite{wang21} between odd- and even-$A$ neighboring nuclei. The HF+BCS equations are then solved iteratively assuming time-reversal and axial symmetry \cite{vau72,vau73}, to obtain the energies, wave functions, and occupation probabilities of the single-nucleon states. The wave functions are expanded in an axially-symmetric harmonic oscillator basis consisting of 11 major shells.
 
From deformation-constrained HF+BCS calculations we obtain energy-deformation curves, {\it i.e.}, the nuclear energy as a function of the nuclear quadrupole deformation parameter $\beta$, defined as
\begin{equation}
\beta = \sqrt{\frac{\pi}{5}}\:\frac{Q_{0}}{A \:\langle r^2 \rangle} \;,
\label{def_param}
\end{equation}
where $Q_{0}$ is the intrinsic nuclear quadrupole moment and $\langle r^2 \rangle$ is the nuclear m.s. radius, both calculated microscopically. Each relative minimum in the energy-deformation curve corresponds to a different equilibrium shape of the nucleus: oblate when $\beta < 0$, spherical when $\beta\approx 0$ and prolate when $\beta > 0$.

The states of odd-$A$ nuclei are built as a one-quasiparticle (1qp) state upon the ground state of the $A-1$ even-even core, the ground state being the 1qp state that minimizes the energy. An occupation probability $v_{\nu}^2=0.5$ is assigned to the single-neutron state that minimizes the total energy of the nucleus, and the remaining probability $v_{\nu}^2=0.5$ is assigned to the corresponding time-reversed state. Time-reversal invariance is thus preserved \cite{rayner2}, this approach being an approximation to the exact blocking when one neglects the fields in the energy density functional that are odd under time reversal \cite{schunck10}.

The m.s. charge radius is given by \cite{neg70,ber72}:
\begin{equation}\label{rch}
\langle r^2_c \rangle = \langle r^2_p \rangle + \langle r^2_c \rangle _p + \frac{N}{Z} \langle r^2_c \rangle _n + r^2_{CM} + r^2_{SO}\, , 
\end{equation}
where $\langle r^2_c \rangle_p = 0.80$ fm$^2$ \cite{sick03} and $\langle r^2_c \rangle_n = -0.12$ fm$^2$ \cite{gentile11} are the proton and neutron m.s. charge radii, respectively (the contribution of the former vanishes in the isotopic differences introduced below), $r^2_{CM}$ and $r^2_{SO}$ are small center-of-mass-motion and spin-orbit corrections, and $\langle r^2_p \rangle$ is the m.s. radius of the density distribution of protons within the nucleus, $\rho_p({\vec r})$, and is given by:
\begin{equation} \label{r2p}
\langle r_p^2 \rangle = \frac{1}{Z} \int r^2 \:\rho_p ({\vec r}) \:d{\vec r} \,.
\end{equation}

It is customary to analyze the evolution of the charge radii in an isotopic chain in terms of their differences with respect to a reference isotope, which in this case is $^{198}$Hg:
\begin{equation}\label{rad_diff}
\delta \langle r^2_c \rangle ^{A,{\rm ref}}= \langle r^2_c \rangle ^A - \langle r^2_c \rangle ^{{\rm ref}} \;.
\end{equation}

Charge radii and their isotopic differences, which can be measured with high precision using laser spectroscopic techniques \cite{cheal}, are very sensitive to the nuclear deformation and reflect the evolution of the nuclear shape in isotopic chains. In fact, the m.s. radius of an axially-deformed nucleus and the spherical value $\langle r^2 \rangle _{\rm sph}$, obtained usually from the droplet model, can be related through an expression that depends on the quadrupole deformation parameter $\beta$ up to second order as
\begin{equation}
\langle r^2 \rangle = \langle r^2 \rangle _{\rm sph} \left(1+\frac{5}{4\pi} \beta^2 \right) \;.
\end{equation}

The magnetic dipole moment of a deformed nucleus contains collective and odd-nucleon contributions and is obtained from the expression \cite{moya86}:
\begin{eqnarray}\nonumber
\mu_J = && g_R \:J + \frac{k^2}{J + 1} \:\Big[g_k - g_R \;+ \Big.\\\label{mag_mom}
&& \Big. + \;\delta_{k,1/2}\: (2J + 1) \:(-1)^{J+1/2} \:\sqrt{2} \:g_{2k} \Big] \;,
\end{eqnarray}
where $g_R$ is a collective gyromagnetic ratio, which is approximated by $Z/A$, and $g_k$, $g_{2k}$ are single-particle (odd-nucleon) gyromagnetic ratios \cite{moya86}. In the states without rotational excitation energy, $J=k$. Recent results on magnetic moments of Hg isotopes from a Gogny Hartree-Fock-Bogoliubov calculation have shown their important role in identifying the ground-state configuration of odd-$A$ nuclei \cite{peru21}.

The GT transitions are obtained as proton-neutron phonon excitations relative to the HF+BCS ground state of the parent nucleus, originated from a residual spin-isospin interaction with particle-hole (ph) and particle-particle (pp) components. The strength of the former can be obtained self-consistently through the second derivative of the energy density functional with respect to the same Skyrme one-body densities that generate the mean field, having an impact on the position and shape of the GT resonance. The latter is a proton-neutron pairing force in the $J^{\pi}=1^+$ coupling channel, whose intensity is fitted to reproduce the general trend of experimental half-lives. These interactions are reduced to a separable form \cite{sarri01npa} with coupling strengths given by $\chi ^{ph}_{GT} = 0.08$ MeV and $\kappa ^{pp}_{GT} = 0.02$ MeV. The sensitivity of the GT strength distributions to the values of these couplings, as well as of the other ingredients of the microscopic calculation such as the parametrization of the Skyrme interaction or the strength and treatment of the pairing correlations, was analyzed in previous works \cite{sarri05,moreno06,boillos15}, and here we make use of the conclusions obtained.

The allowed GT transitions, being simply induced by a spin-isospin operator, do not depend on the radial coordinate, so that transitions between different parent and daughter radial structures (i.e., different shapes) are strongly suppressed. Therefore, the same shape is assumed for the ground state of the parent and for all the GT-populated states in the daughter. This is a good approximation given the small difference between the core deformation caused by an odd neutron or by an odd proton, and is therefore the usual procedure in deformed calculations of the transition amplitudes \cite{moller84}.

At this point, it is appropriate to mention the difficulty of the interpretation of the results in terms of pure shapes when transitional nuclei like those studied here are involved in the decay process. A more elaborated analysis involving a mixing scenario, such as that used in \cite{algora21} might be necessary. In any case, this work provides the basic pieces of information for further developments on those lines, once experimental data become available.

In the GT transitions related to $\beta^+/EC$ decays, the odd neutron in the parent nucleus can act just as a spectator, giving rise to correlated three-quasiparticle (3qp) final states, or it can be involved by pairing up with the new neutron, giving rise to correlated one-quasiparticle (1qp) final states. The transition amplitudes to 3qp final states are calculated with the separable residual interactions mentioned above within a deformed proton-neutron quasiparticle random phase approximation (pnQRPA), resulting in algebraic equations of fourth order in the phonon energy, which are solved numerically. Meanwhile, the transition amplitudes to 1qp final states are obtained by considering phonon correlation perturbations up to first order in the quasiparticle transitions \cite{moe90,muto92,sarri01prc}. These amplitudes are transformed to the laboratory frame by expressing the initial and final states in the basis of intrinsic states through the Bohr-Mottelson factorization \cite{bm}. 

For each type of GT transition, the excitation energy of the final state within the daughter nucleus is:
\begin{eqnarray}
E^*_{(1qp)} &=& E_\pi - E_{\pi_0} \;,\\
E^*_{(3qp)} &=& \omega + E_{\nu_0} - E_{\pi_0} \;,
\label{exc}
\end{eqnarray}
where $E_{\pi}$ is the quasiparticle energy of the odd proton, $\omega$ is the pnQRPA excitation energy, and $E_{\pi_0}$ and $E_{\nu_0}$ are the lowest energies among the proton and neutron quasiparticle states, the latter being also the quasiparticle energy of the odd neutron. A more detailed discussion about the excitation energies in the decay of odd-$A$ nuclei can be found in \cite{moe90}. After a transition to a 3qp final state, the excitation energy is at least twice the nuclear pairing gap, so that the GT strength below typically 2-3 MeV of excitation energy is mainly due to transitions to 1qp final states.

The inverse of the $\beta^+$ or $EC$ half-life is proportional to the sum of the strengths of all GT transitions, $B(GT,E^*)$, corresponding to excitation energies $E^*$ below the $Q_X$ value, where $X$ is $\beta^+$ or $EC$, each transition strength weighted by a phase-space factor $f^X(Z,Q_X-E^*)$:
\begin{eqnarray}\label{t12}
\left[ T^X_{1/2} \right]^{-1} &=& \frac{\left[{\mathcal Q}\:\left( g_{A}/g_{V}\right)\right]^2}{D} \times \\\nonumber
&& \times \sum_{E^*<Q_X} f^X(Z,Q_X-E^*) \:B(GT,E^*) \, ,
\end{eqnarray}
where $D =$ 6144 s is a constant \cite{hardy20} and $(g_A/g_V) = -1.276$ is the ratio between axial and vector charged-current weak couplings. ${\mathcal Q} = 0.77$ is a standard quenching factor for the weak axial coupling \cite{suh17}, which was successfully tested in our previous works against experimental beta-decay data on heavy nuclei using the same theoretical formalism. There are two main types of effects that contribute to the reduction of the weak axial coupling in nuclear processes with respect to its bare (free) value. One is related to the nuclear medium and may involve meson-exchange currents between nucleons and the interference between nucleonic and non-nucleonic degrees of freedom ($\Delta$ excitations). The other is model-dependent and refers to the possible shortcomings of the nuclear-state description due to the theoretical many-body approximations. It is usually very difficult to disentangle the contributions from both types of sources. 

The phase-space factors $f^{\beta^+}$ and $f^{EC}$ are computed taking into account relativistic, finite-nuclear-size, and screening effects \cite{gove71}, and using experimental $Q_{EC}$ values \cite{wang21}. The total $\beta^+/EC$ half-lives are given by:
\begin{eqnarray}
\left[ T^{\beta^+/EC}_{1/2} \right]^{-1} = \left[ T^{\beta^+}_{1/2} \right]^{-1} + \left[ T^{EC}_{1/2} \right]^{-1}
\end{eqnarray}

More details of this theoretical framework can be found in \cite{sarri98,sarri99,sarri01prc,sarri01npa,boillos15}.

\section{Results and discussion}\label{results}

We present in this section the energy-deformation curves of the even-even Hg isotopes from $A=182$ to $A=188$ and identify their equilibrium shapes. We then discuss the possible single-nucleon-state assignments of the odd neutron in the odd-$A$ isotopes in the chain. For each equilibrium shape and spin-parity we show our results of bulk properties (quadrupole deformation parameters, m.s. charge radii, and magnetic dipole moments) and decay properties (GT strength distributions and half-lives).

\subsection{Bulk properties}\label{bulk}

\begin{figure}[ht]
\centering
\includegraphics[width=6.4cm]{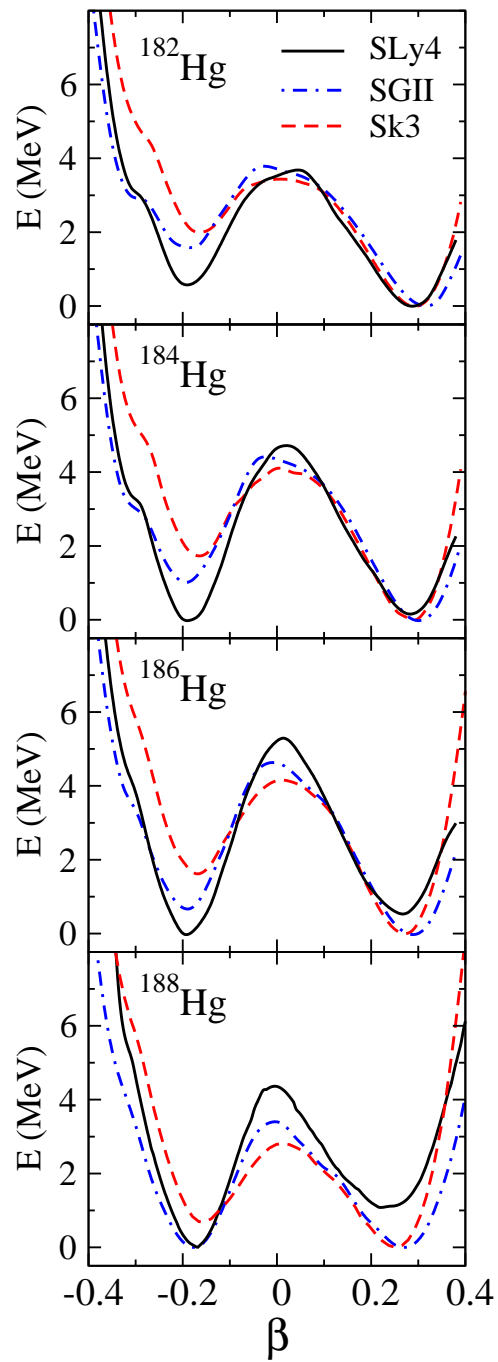}
\caption{ Energy-deformation curves for even-even Hg isotopes from $A=182$ to $A=188$ obtained from a constrained HF+BCS calculation using three parametrizations of the Skyrme interaction: Sk3 (dashed lines), SGII (dashed-dotted lines), and SLy4 (solid lines).}
\label{fig_ener_beta}
\end{figure}

The energy-deformation curves of the even-even Hg isotopes from $A=182$ to $A=188$ obtained from an axially-deformed HF+BCS calculation are shown in Fig.~\ref{fig_ener_beta}, for three parametrizations of the Skyrme interaction (Sk3, SGII, and SLy4). The energies have been shifted in each isotope so that the absolute minima of the curves lie at zero, and the quadrupole deformation parameter $\beta$ is related to the nuclear quadrupole moment as shown in Eq.~(\ref{def_param}). As can be seen in the figure, the three Skyrme forces give similar equilibrium deformations, i.e. the minima of the curves occur approximately at the same values of the deformation parameter $\beta$. The general pattern in this region consists of two equilibrium shapes, one prolate with $\beta$ between 0.25 and 0.30, and one oblate with $\beta$ between $-0.20$ and $-0.15$. 

The relative energies of the minima show a larger dependence on the Skyrme force. For the Sly4 force, the ground state is prolate in $^{182}$Hg and oblate in $^{184}$Hg, $^{186}$Hg, and $^{188}$Hg, with energy differences between both equilibrium shapes of around 1 MeV or lower (almost zero in $^{184}$Hg), and separated by an energy barrier of around 4 MeV centered at $\beta \approx 0$. For the Sk3 and SGII forces the ground state is prolate in all the isotopes, again with small energy differences between both equilibrium shapes, especially in $^{188}$Hg.

The relative energies of the minima and the location of the shape transition in an isotopic chain are indeed sensitive to the details of the calculation, as found here for different Skyrme forces. In other types of calculations the shape transitions take place in different isotopes, for example a beyond-mean-field calculation using the SLy6 Skyrme force \cite{yao13}, where the shape transition occurs between $^{186}$Hg and $^{188}$Hg, or a Gogny D1S \cite{web_gogny} and Gogny D1M \cite{nomura13} mean-field calculation allowing for triaxial deformation, where the shape transition occurs between $^{184}$Hg and $^{186}$Hg. In the latter work a rather flat region was found connecting the two axial minima in the triaxial dimension, whereas in the restricted space of pure axial deformation a potential barrier appears between them, in agreement with our axially-symmetric approach.

Since the nuclear properties of interest in this work depend essentially on the nuclear equilibrium deformation, we will restrict ourselves to just one Skyrme parametrization, the SLy4 force. As described in Section~\ref{theory}, the energy-deformation curves and the location of their minima are very similar in the even-odd and in their neighboring even-even isotopes, due to the small core deformation induced by the odd nucleon. Thus, for the even-even cores of the odd-$A$ Hg isotopes of interest in this work and for the chosen SLy4 Skyrme force, we will consider a prolate equilibrium shape with $\beta\approx0.3$ and an oblate equilibrium shape with $\beta\approx-0.2$.

Experimentally, the isotope $^{183}$Hg has a ground state $1/2^-$ (which has been identified as a rotational band head, b.h.) and excited states $7/2^-$ (b.h.) at an unknown excitation energy (estimated at 120 keV), and $(13/2^+)$ (b.h.) at 183 keV. The isotope $^{185}$Hg has a ground state $1/2^-$ (b.h.) and excited states $(7/2)^-$ (b.h.) at 34 keV, $13/2^+$ (b.h.) at 99.3 keV, and $(9/2^+)$ (b.h.) at 212.7 keV. And the isotope $^{187}$Hg has a ground state $3/2^{(-)}$ and excited states $13/2^{(+)}$ (b.h.) at an unknown excitation energy (possibly 59 keV), and $(9/2^+)$ (b.h.) at 161.6 keV. In our HF+BCS calculation these nuclear states correspond to a 0$^+$ even-even core with a given equilibrium shape, as described in the preceding paragraph, together with an odd neutron in a single-nucleon state near the Fermi level with the experimental spin and parity of the nuclear state. As shown in Table~\ref{table.1}, each deformed single-nucleon state in our calculation is labelled with the quantum numbers $[Nn_z\Lambda]$ that characterize the asymptotic limit of large deformations, which are typically used in the Nilsson model. Our states are indeed linear combinations of the asymptotic states. In our calculation there are several states very close in energy around the Fermi level, and small variations of the nucleon-nucleon interaction can change their relative positions. Therefore, in some cases the odd-nucleon state can be related to more than one Nilsson state.

\begin{table*}[t]
\caption{For the odd-$A$ Hg isotopes $^{183}$Hg, $^{185}$Hg, and $^{187}$Hg, results on magnetic dipole moment $\mu$ in nuclear magnetons, and $\beta^+/EC$ half-life $T_{1/2}^{\beta^+/EC}$ in seconds, from a HF(SLy4)+BCS mean field with spin-isospin correlations in the case of half-lives. We consider different equilibrium shapes, indicated by the quadrupole deformation parameter $\beta$, and odd-neutron states, identified by the corresponding asymptotic quantum numbers in the Nilsson model, which are compatible with the spins and parities $J^{\pi}$ of nuclear states found experimentally. Experimental data of magnetic dipole moments \cite{stone14} and partial $\beta^+/EC$ half-lives \cite{kondev21} are also given.}
\label{table.1}
\begin{tabular}{>{\centering}m{2cm}>{\centering}m{2cm}>{\centering}m{2cm}>{\centering}m{2cm}>{\centering}m{2cm}>{\centering}m{2cm}>{\centering}m{2cm}>{\centering}m{2cm}}\cr
\hline\hline
Isotope & $J^{\pi}$ & $\beta$ & $[Nn_z\Lambda]k^{\pi}$ & $\mu$ $[\mu_N]$ & $\mu_{exp}$ $[\mu_N]$ & $T_{1/2}^{\beta^+/EC}$ [s] & $T_{1/2\;\;exp}^{\beta^+/EC}$ [s] \cr
\hline
\cr
                    & $1/2^{-}$   & $+0.27$   & $[521]1/2^{-}$   & $+0.31$   &+0.524(5)& 22.0 & 10.6(8) \cr
                    &                  & $-0.18$   &  $[521]1/2^{-}$   & $-0.60$  &                   &    9.3 &               \cr
\cr
$^{183}$Hg & $7/2^{-}$   & $+0.27$   & $[514]7/2^{-}$   & $+1.49$   &   --         & 5.9    & --            \cr
                    &                  & $-0.19$  & $[523]7/2^{-}$     & $+0.44$   &              &  11.4 &               \cr
\cr
                    & $13/2^{+}$ & $-0.15$ & $[606]13/2^{+}$  &  $-1.28$  &   --          & 36.2   &                      \cr
\cr
\hline
\cr
                    & $1/2^{-}$   & $+0.25$   & $[521]1/2^{-}$ & $+0.25$  & +0.509(4) & 64.8 & 52.2(12) \cr
                    &                 & $-0.19$     & $[521]1/2^{-}$  & $-0.59$ &               &  17.8    &                \cr
\cr 
                    & $7/2^{-}$   & $+0.27$   & $[514]7/2^{-}$  &  $+1.52$  &  --            & 10.4  & --            \cr
$^{185}$Hg  &                 & $-0.19$   & $[523]7/2^{-}$    &  $+0.42$  &                &  29.3     &                 \cr
\cr
                    & $13/2^{+}$ & $-0.15$ & $[606]13/2^{+}$  & $-1.28$ & $-1.017(9)$ & 86.1 & 47(11)    \cr                  
\cr
                    & $9/2^{+}$   & $+0.27$  &  $[624]9/2^+$   & $-1.01$ &  --            & 73.1 &  --           \cr
                    &                  & $-0.19$    &  $[604]9/2^+$   &  $-0.46$   &               &  51.5 &                \cr
\cr
\hline
\cr
                    & $3/2^{-}$   & $+0.19$   & $[521]3/2^{-}$   &   $-0.34$  & $-0.594(4)$ & 375.3  & 114(18) \cr
                    &                 & $-0.17$  & $[532]3/2^{-}$      &  $+0.38$    &                      & 80.4   &                \cr
\cr                    
$^{187}$Hg & $13/2^{+}$ & $-0.14$ & $[606]13/2^{+}$   &  $-1.28$    & $-1.044(11)$ & 464.2 & 144(18)    \cr
\cr
                    & $9/2^{+}$   & $+0.25$  & $[624]9/2^+$     & $-0.99$    &  --            & 379.8      &  --           \cr
                    &                   & $-0.17$    & $[604]9/2^+$    &  $-0.48$    &               &   155.3      &                \cr
\cr
\hline
\end{tabular}
\end{table*}

Table \ref{table.1} contains the quadrupole deformation $\beta$ obtained for the equilibrium configurations. They agree with the results of other types of calculations based on the Gogny D1S force \cite {web_gogny} as well as with the deformations obtained from macroscopic-microscopic global nuclear calculations of Ref. \cite{moller16}.

\begin{figure}[ht]
\centering
\includegraphics[width=8.5cm]{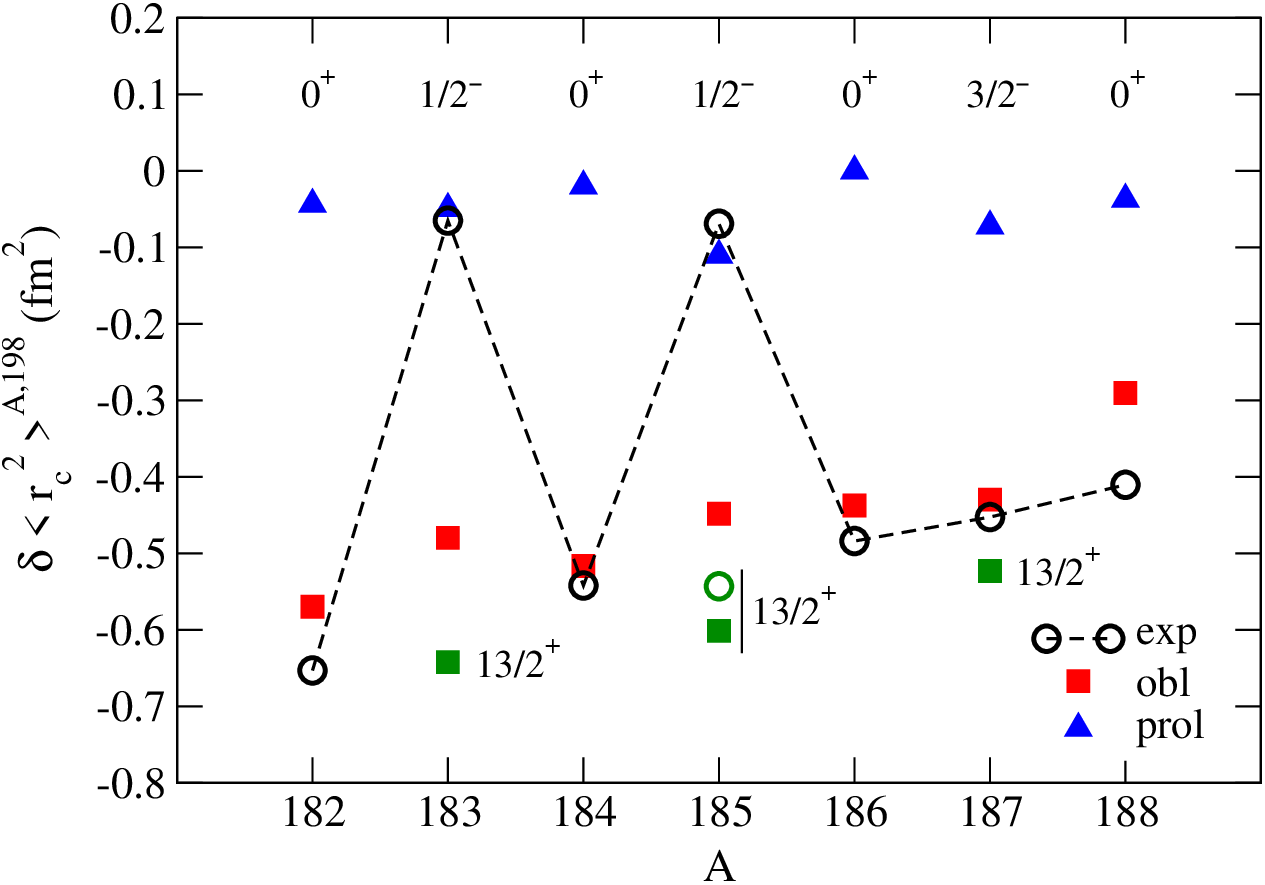}
\caption{ Mean square charge radius differences with respect to $^{198}$Hg , Eq.~(\ref{rad_diff}), from a HF(SLy4)+BCS calculation for Hg isotopes from $A=182$ to $A=188$ in the state with the experimental ground-state spin and parity (given in the figure) for the oblate shape (red squares) and for the prolate shape (blue triangles), together with experimental data \cite{angeli13,marsh18} (black circles joined by a dashed line). For the $13/2^+$ (oblate) isomeric state, the theoretical results of $^{183}$Hg, $^{185}$Hg, and $^{187}$Hg (green squares) and the experimental value of $^{185}$Hg \cite{marsh18} (green circle) are also given.}
\label{fig_r2_sly4}
\end{figure}

The differences of m.s.~charge radii with respect to the reference isotope $^{198}$Hg, Eq.~(\ref{rad_diff}), obtained from a HF(SLy4)+BCS calculation, are shown in Fig.~\ref{fig_r2_sly4} for even- and odd-$A$ Hg isotopes from $A=182$ to $A=188$ in the state with the experimental ground-state spin and parity for each equilibrium shape (oblate and prolate). Experimental data measured with laser spectroscopy techniques \cite{angeli13,marsh18} are also shown. The result for the $13/2^+$ (oblate) isomeric states in $^{183}$Hg, $^{185}$Hg, and $^{187}$Hg are also given, together with the experimental value of $^{185}$Hg \cite{marsh18}.

An even-odd radius staggering is found experimentally \cite{bonn72,ulm86,marsh18} between $^{181}$Hg and $^{186}$Hg, the odd-$A$ isotopes having larger radii. As can be seen in Fig.~\ref{fig_r2_sly4}, our results for an oblate nucleus compare well with the experimental values for even-even Hg isotopes, as well as for $^{187}$Hg. For $^{183}$Hg and $^{185}$Hg the prolate shape results are closer to the experimental values. For Hg isotopes with $A>185$ the prolate shape results remain approximately at the same value, $\delta\langle r^2 \rangle\approx0$, whereas the oblate shape results, compatible with the experimental data, increase smoothly up to $A=190$, where both shapes join and keep increasing with $A$ \cite{boillos15}.

It is worth mentioning another approach based on the Fayans functional, which has been successfully applied to the study of odd-even differential observables in isotopic chains, such as masses or charge radii. This functional incorporates a density-gradient dependence in the pairing interaction to account for a surface tension in the pair condensate \cite{fay00,rein17,bor20}.

In Table~\ref{table.1} we give the values of the magnetic dipole moments in nuclear magnetons ($\mu_N$), Eq.~(\ref{mag_mom}) obtained in our calculation for the different shapes and possible odd-nucleon states compatible with the spins and parities of the low-lying, non-rotational (single-particle) nuclear states found experimentally. For comparison, the magnetic dipole moments at the spherical, extreme independent-particle limit (Schmidt values) are $-1.913$ when the odd neutron is in the $i_{13/2}$, $p_{3/2}$, or $f_{7/2}$ spherical shells, or $+1.565$ when it is in the $h_{9/2}$ shell. We also give in Table~\ref{table.1} the experimental data available \cite{stone14}.

\subsection{Decay properties}\label{decay}

Next, we analyze the decay properties of the isotopes under study, starting with the GT strength distributions for different nuclear equilibrium shapes and odd-neutron states. In previous works, focused on the same \cite{boillos15} and other \cite{sarri05,moreno06} mass regions, it was shown that the theoretical $\beta$-decay patterns are robust against changes in the details of the nucleon-nucleon interaction, which bring about small variations within the distinct profiles obtained for different equilibrium shapes of the same isotope. The sensitivity of the GT strength distributions to the nuclear deformation has been used in past experiments \cite{poi04,nach04,per13,este15,algora21} to extract information on the shape of the decaying nucleus.

\begin{figure}[ht]
\centering
\includegraphics[width=60mm]{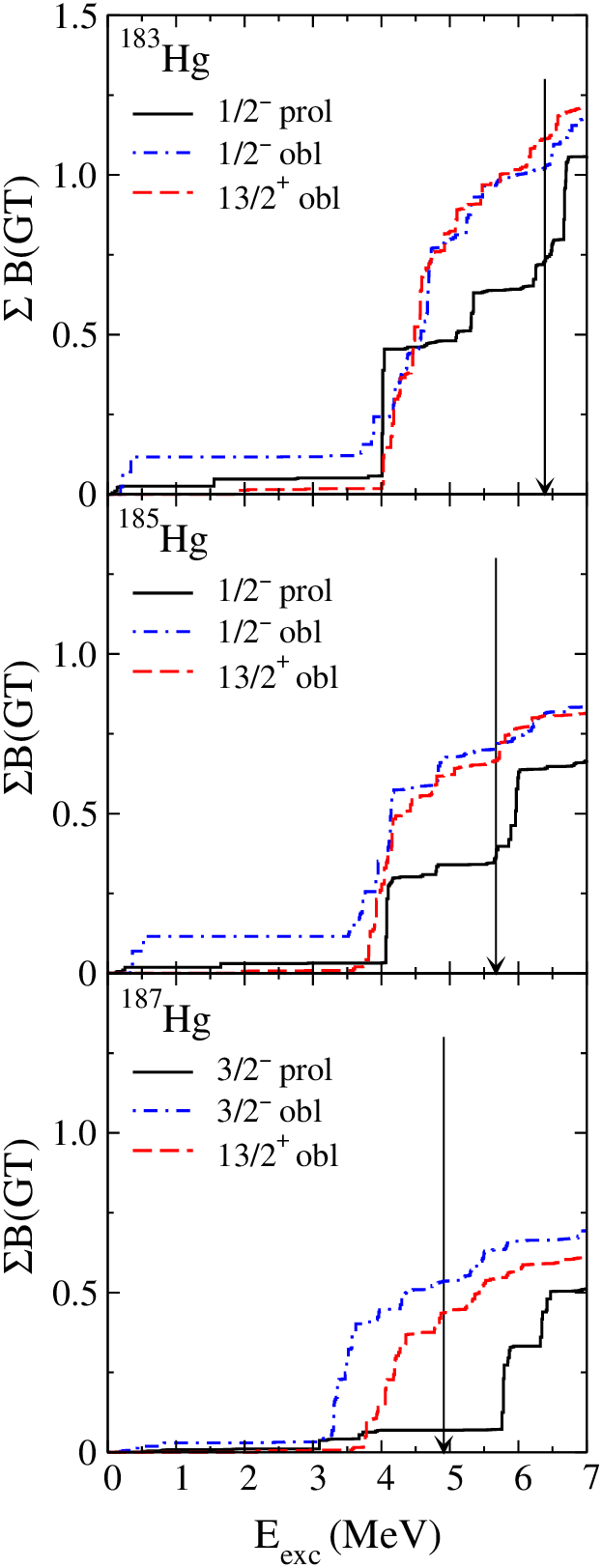}
\caption{ Accumulated GT strength distributions in $g_A^2/4\pi$ units for $^{183}$Hg, $^{185}$Hg, and $^{187}$Hg from a HF(SLy4)+BCS+pnQRPA(ph,pp) calculation. The 1/2$^-$ or 3/2$^-$ ground states are shown for two equilibrium shapes (solid black line for the one compatible with the experimental m.s. charge radius, dash-dotted blue line for the other shape), and the 13/2$^+$ isomeric state is also given (red dashed line). The experimental $Q_{EC}$ value \cite{wang21} is indicated by a vertical arrow.}
\label{fig_bgt_odd_sly4}
\end{figure}

Fig.~\ref{fig_bgt_odd_sly4} shows the main results of this work, focused on a future ISOLDE-CERN measurement \cite{is707} aimed at the independent measurement of the $\beta$-decay patterns of the ground and isomeric states from isomerically purified odd-$A$ Hg beams. The plot shows the accumulated GT$^+$ strength distributions of the ground states of $^{183}$Hg, $^{185}$Hg, and $^{187}$Hg (with spins and parities $1/2^-$, $1/2^-$, and $3/2^-$, respectively) for the two equilibrium shapes given in Table~\ref{table.1} (oblate and prolate), together with the result for the isomeric $13/2^+$ state (oblate). The experimental $Q_{EC}$ value of each isotope \cite{wang21} is indicated by a vertical arrow (6.387 MeV for $^{183}$Hg, 5.674 MeV for $^{185}$Hg, and 4.910 MeV for $^{187}$Hg).

The strength, given in $g_A^2/4\pi$ units, has been obtained from a HF(SLy4)+BCS+pnQRPA(ph,pp) calculation. In these plots the GT strengths are calculated purely in terms of the nuclear matrix elements and do not include the axial-coupling quenching factor (as appeared in Eq.~(\ref{t12})). However, in order to compare with the $\beta$-strength distributions extracted from future experiments, one should include an overall reduction given by the quenching factor squared, ${\mathcal Q}^2\approx0.6$. This value is the same for the whole energy range of the GT strength distribution, so that the richness of detail and the signatures of deformation, spin and parity of the decaying nucleus in the GT pattern, which are the main focus of this work, are essentially unaffected.

If we compare the GT strength distribution corresponding to the experimental ground-state spin-parity assignments but to different deformations (solid black and dash-dotted blue lines), we can see that the oblate shapes of $^{183}$Hg and $^{185}$Hg, with $1/2^-$ ground states, generate strength at very low excitation energy below 1.5 MeV, which is practically absent in the prolate case. In $^{187}$Hg, with $3/2^-$ ground state, both shapes become clearly distinct at around 3 MeV. For $^{183}$Hg and $^{185}$Hg the prolate configurations produce sharp peaks of strength around 4 and 6 MeV, whereas the strength for oblate configurations is more fragmented, continuously accumulating strength in that interval, but especially between 4 and 5 MeV. In the case of $^{187}$Hg, the prolate shape produces practically no strength below $Q_{EC}$, while the oblate shape accumulates a significant amount of strength between 3 and 4 MeV. This could be a valid signature for the type of deformation. In the case of the isomeric state $13/2^+$, there is basically no strength below 4 MeV followed by a significant peak around 4 MeV, especially in $^{183}$Hg and $^{185}$Hg. A total absorption spectroscopy measurement \cite{is707} should be sensitive to these features.

\begin{figure}[ht]
\centering
\includegraphics[width=60mm]{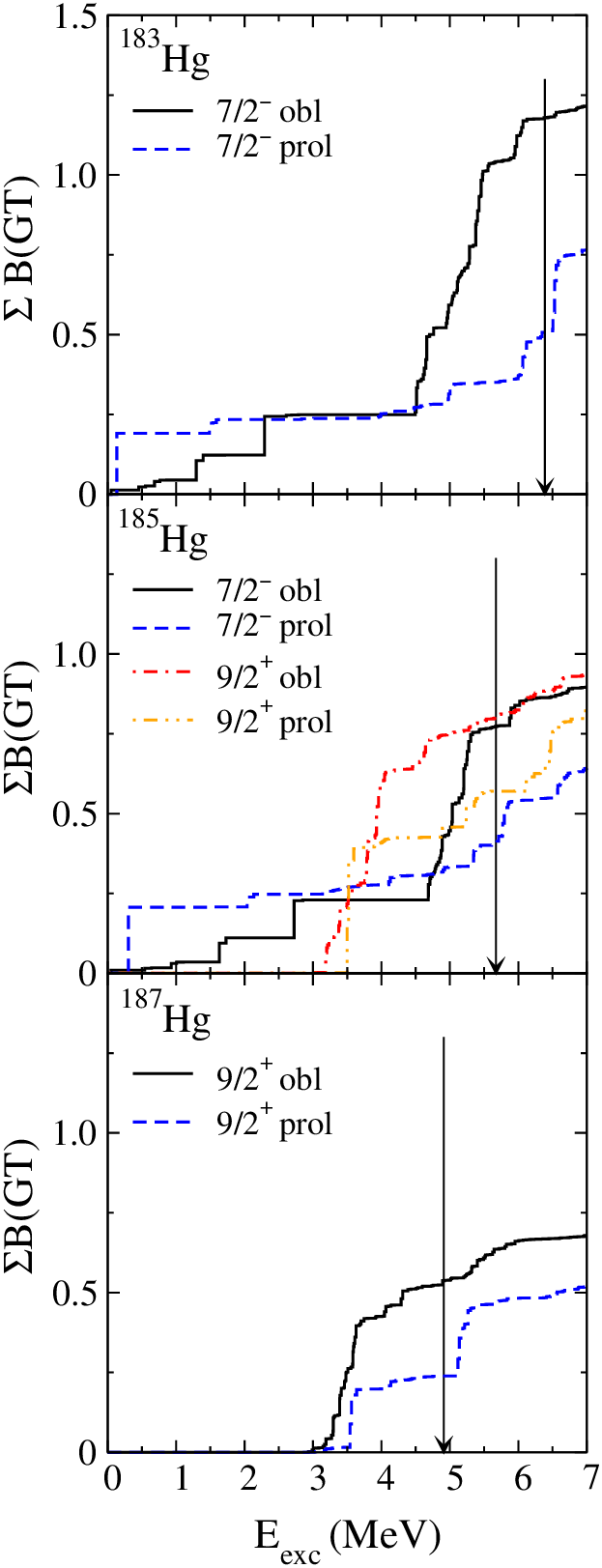}
\caption{ Same as in Fig. \ref{fig_bgt_odd_sly4} but for additional excited states found experimentally in $^{183}$Hg, $^{185}$Hg, and $^{187}$Hg, with two equilibrium shapes for each of them.}
\label{fig_bgt_odd_sly4_other_states}
\end{figure}

In Fig.~\ref{fig_bgt_odd_sly4_other_states} we show the results for other excited states identified experimentally: $7/2^-$ for $^{183}$Hg and $^{185}$Hg, and $9/2^+$ for $^{185}$Hg and $^{187}$Hg. In each case the two equilibrium shapes given in Table~\ref{table.1} (oblate and prolate) are considered.

As the number of neutrons increases in this Hg isotopic chain, approaching stability, a general trend of decreasing GT strength is seen in the figure. Another general feature is that the strength is more fragmented in the oblate shapes than in the prolate shapes, so that the accumulated strength increases more steadily in the former. This may provide an additional signature for the type of deformation and could make the proposed experiment \cite{is707} with purified decaying states sensitive to the nuclear shape.

One expects an influence of the spin-parity assignment of the odd-neutron state on the GT strength distributions, especially at low energies, where the odd neutron is involved in the transitions to 1qp final states. However, the influence of spin and parity within the same equilibrium shape is usually smaller than the influence of the deformation itself, as was discussed in the description of Fig.~\ref{fig_bgt_odd_sly4}, where the distributions for the oblate shapes are close to each other, although they correspond to different spin-parity (1/2$^{-}$ vs. 13/2$^{+}$ in $^{183,185}$Hg, 3/2$^{-}$ vs. 13/2$^{+}$ in $^{187}$Hg), whereas they substantially differ from the profiles for the prolate shapes, even when they correspond to the same spin-parity (1/2$^{-}$ in $^{183,185}$Hg, 3/2$^{-}$ in $^{187}$Hg). In other words, typically the GT strength distributions contain signatures of the nuclear deformation that remain when different odd-neutron spin-parity assignments are considered. 

\begin{figure}[ht]
\centering
\includegraphics[width=60mm]{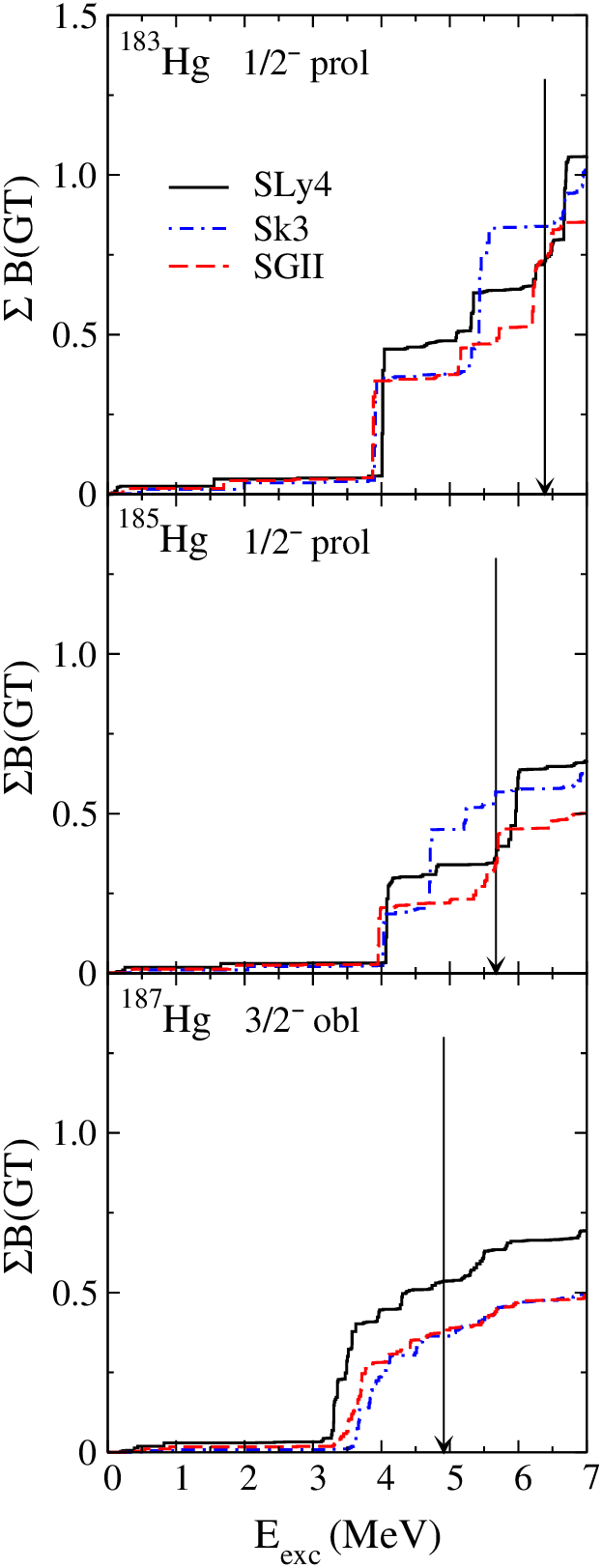}
\caption{ Same as in Fig. \ref{fig_bgt_odd_sly4} but for different Skyrme force parametrizations: SLy4 (solid black line), Sk3 (dash-dotted blue line), and SGII (dashed red line), for the theoretical ground states in $^{183}$Hg, $^{185}$Hg, and $^{187}$Hg compatible with the experimental m.s. charge radius.}
\label{fig_bgt_odd_forces}
\end{figure}

The sensitivity of the GT strength distributions against changes in the nucleon-nucleon interaction is illustrated in Fig.~\ref{fig_bgt_odd_forces} for the theoretical ground states of $^{183}$Hg, $^{185}$Hg, and $^{187}$Hg compatible with the experimental m.s. charge radius, namely $1/2^-$ prolate in $^{183}$Hg and $^{185}$Hg and $3/2^-$ oblate in $^{187}$Hg. Three Skyrme parametrizations have been used: SLy4 (the choice in the previous figures), Sk3, and SGII. As a general feature, the three forces produce important accumulations of GT strength around 4, 5 and 6 MeV of excitation energy in $^{183}$Hg, around 4 and 6 MeV in $^{185}$Hg, and between 3 and 4 MeV in $^{187}$Hg. The Sk3 force in $^{185}$Hg differs from the others in that it gives rise to a significant concentration of strength slightly below 5 MeV instead of at 6 MeV.

\begin{figure}[ht]
\centering
\includegraphics[width=50mm]{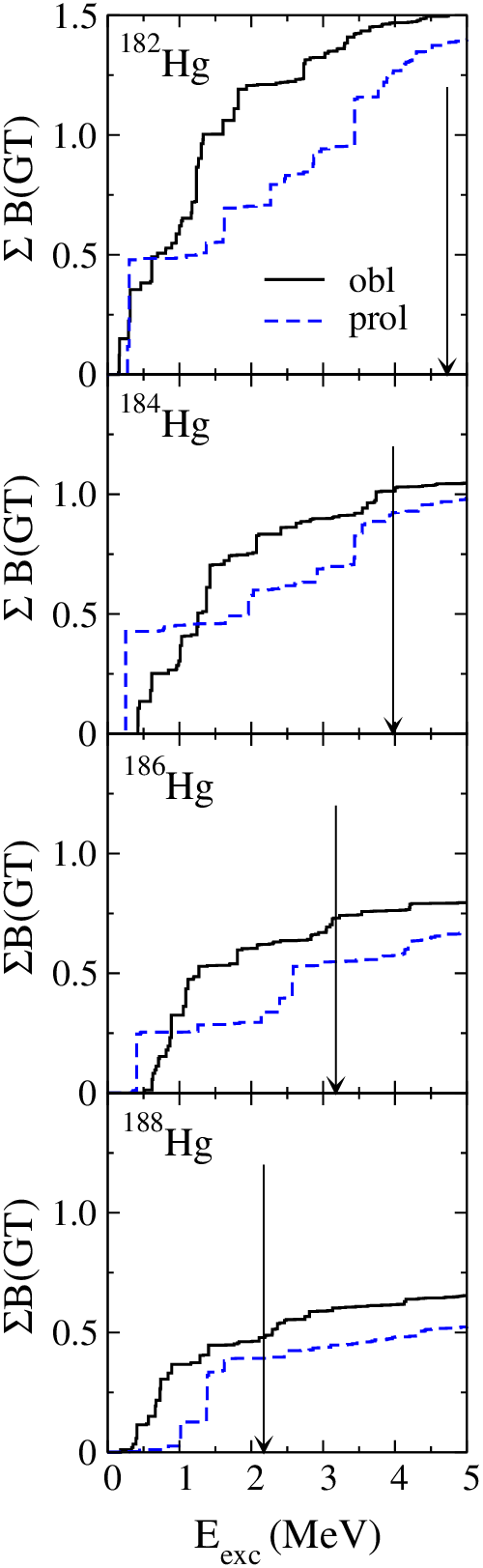}
\caption{ Same as in Fig. \ref{fig_bgt_odd_sly4} but for the ground states of the neighboring even-even isotopes: $^{182}$Hg, $^{184}$Hg, $^{186}$Hg, and $^{188}$Hg for two equilibrium shapes: oblate (solid black line) and prolate (dashed blue line).}
\label{fig_bgt_even_sly4}
\end{figure}

Finally, we complement the study of the decay properties in this region with the accumulated GT strength distributions of the neighboring even-even Hg isotopes, shown in Fig.~\ref{fig_bgt_even_sly4}. This strength is shifted to lower excitation energies in comparison with the odd-$A$ isotopes, since in the latter the transitions to 3qp final states occur typically above twice the pairing gap (2-3 MeV). The $Q_{EC}$ values are also reduced by a similar amount in the even-even isotopes with respect to the odd-$A$ ones.

As described in Sec. \ref{theory}, the half-life of a $\beta^+/EC$ decay, Eq.~(\ref{t12}), depends on the strength of the GT$^+$ transitions to accesible excited states in the daughter nucleus, namely those with excitation energy below the $Q_{EC}$ value. The accumulated strength at that point, which is directly observable in experiments, increases as one moves away from the stability towards lighter, more neutron-deficient isotopes.

In the low-energy region, particularly within the $Q_{EC}$ window, the accumulated GT strength for the oblate shapes are generally larger than for the prolate shapes. As discussed above, the spin-parity assignment of the odd neutron is also expected to have an influence on the GT strength within the $Q_{EC}$ window, where the transitions to 1qp final states, which involve the odd neutron, are predominant.

\begin{figure}[ht]
\centering
\includegraphics[width=7cm]{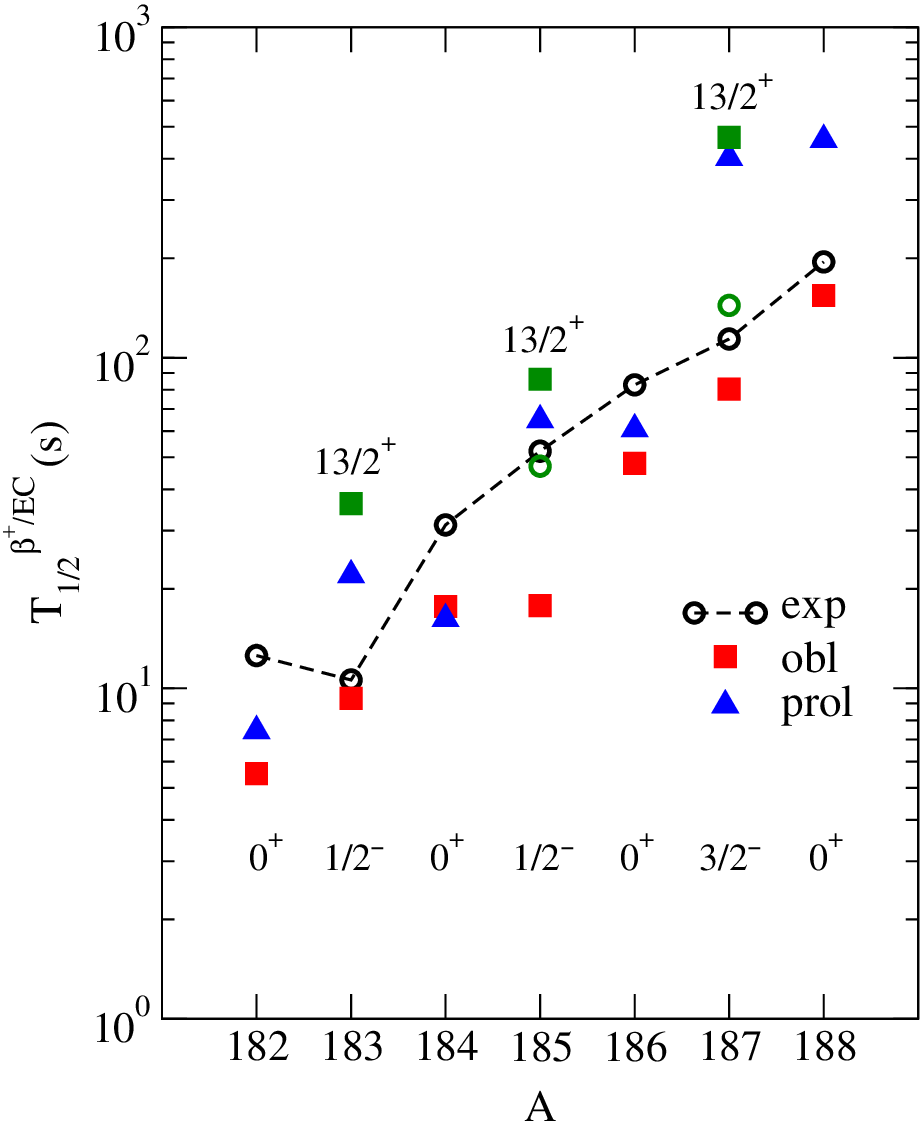}
\caption{ Half-lives for the $\beta^+/EC$ decay of $^{183}$Hg, $^{185}$Hg, and $^{187}$Hg, from a HF(SLy4)+BCS+pnQRPA(ph,pp) calculation in the state with the experimental ground-state spin and parity for the oblate shape (red squares) and for the prolate shape (blue triangles), together with experimental data \cite{kondev21} (black circles joined by a dashed line). For the $13/2^+$ (oblate) isomeric state, the theoretical results of $^{183}$Hg, $^{185}$Hg, and $^{187}$Hg (green squares) and the experimental value of $^{185}$Hg and $^{187}$Hg \cite{kondev21} (green circle) are also given.}
\label{fig_t12_sly4}
\end{figure}

The results for the $\beta^+/EC$ half-lives from a HF(SLy4)+BCS+pnQRPA(ph,pp) calculation are given in Table~\ref{table.1} for different equilibrium shapes and odd-nucleon states. In Fig.~\ref{fig_t12_sly4} we plot the theoretical $\beta^+/EC$ half-lives of the Hg isotopic chain with $A$ from 182 to 188, including even- and odd-$A$ isotopes. In both Table~\ref{table.1} and Fig.~\ref{fig_t12_sly4} we also include the partial $\beta^+/EC$ half-lives extracted from experimental data on total half-lives and $\beta^+/EC$ branching ratios \cite{kondev21}. For the odd-$A$ isotopes, results are shown for the state with the experimental ground-state spin and parity for both equilibrium shapes, as in Fig.~\ref{fig_r2_sly4}.

In the two heaviest isotopes considered, $^{188}$Hg and $^{187}$Hg, the half-lives of the oblate shapes compare better with the experimental values. In contrast, in $^{186}$Hg and $^{185}$Hg it is the prolate shape that gives better results. In $^{184}$Hg both shapes differ equally from the experimental value, in $^{183}$Hg the oblate shape clearly gets closer, and in $^{182}$Hg the prolate shape is slightly better. A similar trend is found when using the Sk3 and SGII Skyrme forces. One should keep in mind that the half-life condenses in just one number a rich pattern of GT strength and that it is very sensitive to the details of the force generating the mean field and, more particularly, the spin-isospin correlations. In addition, there is a pronounced dispersion of results for different odd-neutron states within the same nuclear shape, as can be seen in Table~\ref{table.1}. In view of these considerations, the half-lives by themselves do not seem to represent a distinctive signature of the shape of the decaying nucleus. Nonetheless, it should be pointed out that the decay half-lives may be directly measured in the proposed experiment \cite{is707} with isomerically purified $\beta$-decay states.

\section{CONCLUSIONS}\label{conclusions}

We have focused on the analysis of bulk and decay properties of the odd-$A$ nuclei $^{183}$Hg, $^{185}$Hg and $^{187}$Hg, obtained from a self-consistent, axially-deformed HF mean-field calculation with a Skyrme SLy4 effective nucleon-nucleon interaction and fixed-energy-gap pairing correlations within BCS approximation, and including spin-isospin residual interactions in separable form in the pp and ph channels within a pnQRPA formalism. The calculation gives rise to two equilibrium shapes for the isotopes under study, one oblate ($\beta\approx-0.2$) and one prolate ($\beta\approx0.3$), in agreement with other theoretical works based on Gogny mean-field, relativistic mean-field, or Monte Carlo shell model approaches, among others. The states of odd-$A$ nuclei have been obtained by locating the odd neutron in the single-nucleon states near the Fermi level of the even-even core. Some single-nucleon states are very close in energy and therefore different possibilities of the odd-neutron location have been considered for the ground or excited nuclear states. For each equilibrium shape and selected odd-neutron states compatible with the experimental spins and parities of low-lying nuclear states, we have computed bulk properties such as quadrupole deformations, charge radius differences, and magnetic dipole moments, and we have compared them to the available experimental data.

An even-odd radius staggering is found in the isotopic chain between $^{181}$Hg and $^{188}$Hg, in agreement with experimental data, where the odd-$A$ isotopes $^{183}$Hg and $^{185}$Hg have larger radii than their even-$A$ neighbors. This effect of the radii can be related to an alternation of the nuclear equilibrium shape. The analyzed ground-state bulk properties are in general compatible with a prolate shape in the odd-$A$ isotopes and with an oblate shape in the even-$A$ isotopes. From $^{186}$Hg onwards, the radii staggering fades away and the ground states are all compatible with an oblate shape.

Next we have analyzed the decay properties, in particular the GT$^+$ strength distributions, whose measurement is planned at ISOLDE-CERN \cite{is707} using TAGS techniques. These results have been obtained by introducing spin-isospin correlations within pnQRPA for different equilibrium shapes and odd-nucleon states, including those compatible with the experimental ground and isomeric nuclear states. We have also given the corresponding $\beta^+/EC$ half-lives and compared them with experimental data, although the theoretical uncertainties of this number make the interpretation more ambiguous. In contrast, the richness of the GT patterns allows for a clearer identification of signatures of deformation and of spin and parity in the decaying nuclei. This analysis was based on the assumption of pure shapes for parent and daughter nuclei. Comparison of these results with future experimental data on these Hg isotopes will help determine the amount of mixing between the different shape configurations in the decay partners.

The predictions presented here are aimed at providing a theoretical guidance to the future measurements of GT strength in odd-$A$ nuclei separately for the ground state and for isomeric states, which will allow for the spin and parity selection of the decaying state and therefore for the disentanglement between the decay patterns of oblate and prolate shapes in the same nucleus.

\begin{acknowledgments}
This work was supported by Ministerio de Ciencia e Innovaci\'on (Spain) under Contract Nos.~PGC2018-093636-B-I00 (P.S. and O.M.), RTI2018-098868-B-I00 (O.M. and L.M.F.), and PID2019-104714GB-C21 (A.A. and S.E.A.O.), by Grupo de F\'isica Nuclear (GFN) at UCM (O.M. and L.M.F.), by the National Research, Development and Innovation Fund of Hungary, financed under the K18 funding scheme with Projects Nos.~K~128729 and NN128072 (A.A.), and by the Generalitat Valenciana Grant No.~PROMETEO/2019/007 (S.E.A.O.).
\end{acknowledgments}


\end{document}